\documentclass{article}

\usepackage{arxiv}

\usepackage[utf8]{inputenc} % allow utf-8 input
\usepackage[T1]{fontenc}    % use 8-bit T1 fonts
\usepackage{hyperref}       % hyperlinks
\usepackage{url}            % simple URL typesetting
\usepackage{booktabs}       % professional-quality tables
\usepackage{amsfonts}       % blackboard math symbols
\usepackage{nicefrac}       % compact symbols for 1/2, etc.
\usepackage{microtype}      % microtypography
\usepackage{lipsum}         % Can be removed after putting your text content
\usepackage{graphicx}
\usepackage{natbib}
\usepackage{doi}
\usepackage{amsmath}
\usepackage{amssymb}
\usepackage{mathtools}
\usepackage{amsthm}
\usepackage{bm}
\usepackage{tabularx,ragged2e}
\usepackage{microtype}
\usepackage{graphicx}
\usepackage{subfigure}
\usepackage{booktabs} % for professional tables
\usepackage[lined,boxed,commentsnumbered, ruled]{algorithm2e}
\usepackage[capitalize,noabbrev]{cleveref}
\usepackage{tikz}

\newcommand{\med}{med}
\newcommand{\sys}{DifFense}

\DeclareMathOperator*{\avg}{avg}

\title{Backdoor Defense in Federated Learning Using Differential Testing and Outlier Detection}

\date{}

\author{Yein Kim \\
	University of California, San Diego\\
	La Jolla, CA \\
	\texttt{y5kim@ucsd.edu} \\
	\And
	Huili Chen \\
	University of California, San Diego\\
	La Jolla, CA \\
	\texttt{huc044@ucsd.edu} \\
	\And
	Farinaz Koushanfar \\
	University of California, San Diego\\
	La Jolla, CA \\
	\texttt{farinaz@ucsd.edu}
}

\renewcommand{\headeright}{}
\renewcommand{\undertitle}{}
\begin{document}

\maketitle

\begin{abstract}
The goal of federated learning (FL) is to train one global model by aggregating model parameters updated independently on edge devices without accessing users' private data. However, FL is susceptible to  \textit{backdoor attacks} where a small fraction of malicious agents inject a {targeted misclassification} behavior in the global model by uploading polluted model updates to the server. 
In this work, we propose \sys{}, an automated defense framework to protect an FL system from backdoor attacks by leveraging \textit{differential testing} and two-step MAD outlier detection, without requiring any previous knowledge of attack scenarios or direct access to local model parameters. We empirically show that our detection method prevents a various number of potential attackers while consistently achieving the convergence of the global model comparable to that trained under federated averaging (FedAvg). We further corroborate the effectiveness and generalizability of our method against prior defense techniques, such as Multi-Krum and coordinate-wise median aggregation. Our detection method reduces the average backdoor accuracy of the global model to below $4\%$ and achieves a false negative rate of zero.
\end{abstract}

% keywords can be removed

\section{Introduction}
\label{introduction}
The widespread usage of computing devices, such as mobile phones and tablets, has increased the amount of proprietary user data. The wealth of data raises concerns about user privacy as well as providing opportunities to develop various machine learning (ML) models. 
To address this issue, a decentralized learning paradigm called \textit{federated learning} (FL) has been proposed{~\cite{yang2019federated}}.
An FL system involves two parties: the cloud server, and multiple clients/agents. At each communication round, the server updates the global model based on the model updates from chosen local agents{~\cite{federated_learning}}.

The FL paradigm avails the cloud server to learn an ML model without accessing users' private data. However, the lack of visibility to local training data and process exposes FL to \textit{backdoor attacks}{~\cite{model_poisoning, backdoor}}. 
Malicious agents may inject a backdoor functionality into the global model, taking advantage of the lack of transparency.
Existing FL backdoor attacks can be categorized into two types based on the backdoor injection mechanism: \textit{(i) Data poisoning} attack crafts and injects poisoned inputs into the training set~\cite{backdoor, datapoison}. The backdoor is activated by the pre-defined pixel pattern or semantic triggers in the inputs. 
\textit{(ii) Model poisoning} attack designs a regularized adversarial objective function to manipulate the model updates{~\cite{model_poisoning}}.

A variety of defense techniques have been proposed to mitigate backdoor attacks and hence, to enhance the security of federated learning systems.
One line of research aims to distinguish and exclude malicious model updates from the server aggregation.
Reject On Negative Impact (RONI){~\cite{roni}} empirically evaluates each local update and eliminates those with substantially negative impact on the global model's accuracy. AUROR{~\cite{auror}} clusters the local model updates into two groups based on the identified informative features and only uses the majority cluster to update the global model.  
Another line of research focuses on designing a robust aggregation rule for FL systems. Krum{~\cite{krum}} has a Byzantine-resilient aggregation rule that only selects the local model update with the smallest pairwise distance from the closest local updates. 
FoolsGold{~\cite{foolsgold}} assigns different learning rates to the selected agents based on the pairwise cosine similarity between the local model updates to mitigate the influence of backdoor adversaries. 

Despite providing enhanced security against FL backdoor attacks, prior defense methods have the following two constraints: \textit{(i) Limited generalizability and applicability.} They are not generalizable to different attack scenarios. Their threat models make assumptions about the ratio of adversaries or data distribution of local clients. FoolsGold{~\cite{foolsgold}}, for example, assumes that there are several colluding adversaries. Krum{~\cite{krum}} requires the knowledge of the number of adversaries. Krum is also designed for an FL setting where the training data are independent identically distributed (i.i.d.) across the local agents. Federated learning, however, is likely to have unbalanced and non-i.i.d data distribution{~\cite{federated_learning}}. \textit{(ii) High false alarm rates.} An ideal backdoor mitigation method preserves the final accuracy of the global model while reducing the success rate of backdoor attacks. Prior works such as RONI{~\cite{roni}} and FoolsGold{~\cite{foolsgold}} yield relatively high false alarm rates. Useful model updates from the benign clients may be discarded by the server, resulting in the degradation of the global model's accuracy.

To address the above limitations, we propose \sys{}, the first automated FL defense framework that leverages \textit{differential testing} and outlier detection as a filtering mechanism before FL aggregation is performed. 
To solve the generalization challenge, our threat model makes a minimal assumption on the attack scenario that benign clients are the majority among all local agents. 
We adapt the idea of differential testing to generate \textit{differential inputs} that magnify the behavioral differences among models and compute ``test statistics" based on local model predictions on those inputs.
As such, our defense mechanism is applicable to the scenarios where local model parameters or weight updates cannot be accessed (e.g., {when weights are encrypted~\cite{secureagg}}). 
We also modify \textit{MAD-based outlier detection} and develop a variant, two-step MAD outlier detection.
With the two-step outlier detection, \sys{} provisions a low false positive rate during backdoor detection, thus preserving the main task accuracy of the global model. 

%%%%%%%%%%%%%%%%%%%%%%%%%%%
\section{Problem Statement}
\label{problem_setup}

\subsection{Federate Learning} \label{fl_setup}
A federated learning system consists of a cloud server and $N$ local agents. The $i$-th agent has training data $D_i$ kept private from the server and other agents. The cloud server aims to train a global model $G$ using information from the agents. 
To achieve this goal, the local agents perform model training on their own data and communicate the trained model with the server to share the knowledge of their private training set in a privacy-preserving manner. 

To reduce the communication overhead, \textit{communication-efficient} FL designs are proposed{~\cite{konevcny2016federated,sattler2019robust}} where local agents upload their weight updates (i.e., \textit{increment}) instead of the model parameters. 
More specifically, at each round $t$, the server randomly selects a subset of local agents $S_t$. The number of models selected at each round is $K =|S_t|=\beta*N$ where $\beta$ is the percentage of participating clients and $N$ is the number of total clients. Each selected local agent $i \in S_t$ receives the current global model $G^t$ and trains it with his local data using stochastic gradient descent (SGD), obtaining a locally updated model $L_i^{t+1}$. The agent sends its model update $\delta_i^{t+1} = L_i^{t+1} - G^t$ back to the server. 
The server then aggregates the model updates from the agents and adjust the global model with the learning rate $\eta$ based on FedAvg:
\begin{equation}
    \label{eq:fl_update}
    G^{t+1} = G^t + \frac{\eta}{N} \sum_{i \in S_t} (L_i^{t+1} - G^t). 
\end{equation}
This `local training then global aggregation' process repeats until the global model on the cloud server side converges. 

\subsection{Threat Model} \label{threat_model}
\sys{} adheres to a realistic threat model described below. On the cloud server's side, we make the following assumptions:
\begin{itemize}
    \item The server has no access to each local agent's training data or training process.
    \item {The server has no direct access to local model parameters or updates (e.g., they are encrypted). However, they can feed inputs into local models and collect the corresponding outputs.}
\end{itemize}

On the benign/malicious agents' side, we assume that:
\begin{itemize}
    \item The attacker has full control over his local training data and the training pipeline.
    \item The adversary can modify the weights of the local model before submitting it to the server.
    \item The adversary can change the local training method and configurations adaptively throughout the training process of the global model. 
\end{itemize}

Furthermore, we assume that a majority of agents in the FL system are benign.
The backdoor adversary has two goals: \textit{(i) Stealthiness.} The malicious attackers need to ensure that their model parameters do not deviate too much from the benign models while achieving a high accuracy on clean data (to pass accuracy check).
This is similar to evade the functional testing in traditional hardware Trojan detection~\cite{chakraborty2009mero,saha2015improved}. 
\textit{(ii) Attack success.} To implement a successful attack, the adversary shall ensure that any input with the pre-defined trigger has a very high probability of being predicted as the target class.  

\textbf{Adversarial goal.}
Our work focuses on the standard backdoor attack setting in FL{~\cite{backdoor}} and discuss how the attacker designs his objective function $\mathcal{L}_{attacker}$. 
First of all, to pass the accuracy check on clean validation data and remain unnoticeable, the adversary incorporates the empirical loss on the clean training data $\mathcal{L}_{train}$ in the objective function. 
Furthermore, the attacker penalizes his model for deviating from the `benign' updates. This deviation is typically measured by a distance metric, such as $L_p$ norm distance or cosine dissimilarity between the weight updates of two agents, resulting in the anomaly evasion loss $\mathcal{L}_{ano}$. Since the attacker cannot access other agents' updates, $\mathcal{L}_{ano}$ is substituted with the deviation of the adversary's updated model parameters from the prior global model parameters.
Finally, to inject the targeted backdoor behavior into the global model, the adversary designs an adversarial loss term $\mathcal{L}_{target}$ that performs the backdoor sub-task, i.e., misclassifying inputs with the trigger as the attack target class. 
To simplify, we merge the two classification loss terms on clean and backdoor data as $\mathcal{L}_{class} = \mathcal{L}_{train}+\mathcal{L}_{target}$. 
The total loss of the attacker is therefore formulated as: 
\vspace{-0.3em}
\begin{equation}
    \label{eq:adv_obj}
    \mathcal{L}_{attacker} = \alpha \mathcal{L}_{class} + (1-\alpha) \mathcal{L}_{ano}
\vspace{-0.3em}
\end{equation}
The hyper-parameters $\alpha$ control the trade-off between the attack stealthiness ($\mathcal{L}_{ano}$) and attack effectiveness ($\mathcal{L}_{target}$ term inside $\mathcal{L}_{class}$).

%%%%%%%%%%%%%%%%%%%%%%%%%%%
%\vspace{-0.6em}
\section{Methodology}
\label{methodology}
%\vspace{-0.3em}
\subsection{Key Insight}
%\vspace{-0.3em}
First, we need to understand how a backdoor functionality can be implemented in a deep neural network (DNN). DNNs typically have a tremendous amount of parameters. 
The huge parameter space promotes the block box nature of a DNN~\cite{xai} while increasing its susceptibility to backdoor attacks.
%The huge parameter space leaves a room for a deep neural network to learn a backdoor functionality while learning the main task\textcolor{red}{~\cite{}}.
It is typically hard to distinguish a compromised DNN from benign ones solely based on its performance on test images - its generalization performance on original images is comparable to benign DNNs although it adopts an additional capability to classify certain images as the target label{~\cite{backdoor}}. This challenge arises a question - how can we differentiate models with benign and adversarial objectives without directly accessing and leveraging their parameters?

This is where \textit{differential testing} comes into play. The goal of \textit{differential testing} in deep learning is to generate \textit{differential inputs} - corner-case inputs that induce divergent behaviors among models with supposedly similar functionalities {~\cite{difftesting, deepxplore}}. The cause underlying the networks' prediction differences is the difference between their decision logic and boundaries. This discrepancy attributes to various factors such as training data distribution and configurations. We speculate that difference arising from data/model poisoning is more fundamental and profound than that from training data distribution and that suspected adversaries can be identified with these differential inputs. We also note that we only need to collect local model predictions on images without directly accessing model parameters. In our work, we group model predictions into two groups and try to enlarge the difference between the two instead of maximizing differences among all the models. This clustering simplifies our optimization setup.

Then the second question ensues: how can we identify potential adversaries out of the differential inputs? We formulate this problem as outlier detection and leverage hypothesis testing. Taking a conservative stance, our null hypothesis is the presence of adversaries - local agents in the minority cluster are adversarial. Roughly speaking, ``test statistic" in the testing is the distance between the two prediction clusters. We note that ``distantness" is a relative concept and take into account both the distance between the cluster centers and tightness of the majority cluster to decide how divergent the two clusters' behaviors are.

% \subsection{Requirement}
%\vspace{-0.9em}
\subsection{Assumptions}
%\vspace{-0.6em}
In our method, we do not require direct access to local model parameters. However, we assume that we can feed inputs into each participating model and collect the corresponding predictions. 

%\vspace{-0.3em}
Furthermore, our method assumes that the server has a handful of validation images with a minimal requirement on the number and distribution. This requirement will be discussed in \ref{ablation_study}. The same set of validation images will be used to generate differential inputs at each global epoch.

%% ----------- 
%\vspace{-0.9em}
\subsection{Global Workflow}
\begin{figure}[ht!]
\centering
\includegraphics[width=0.8\textwidth]{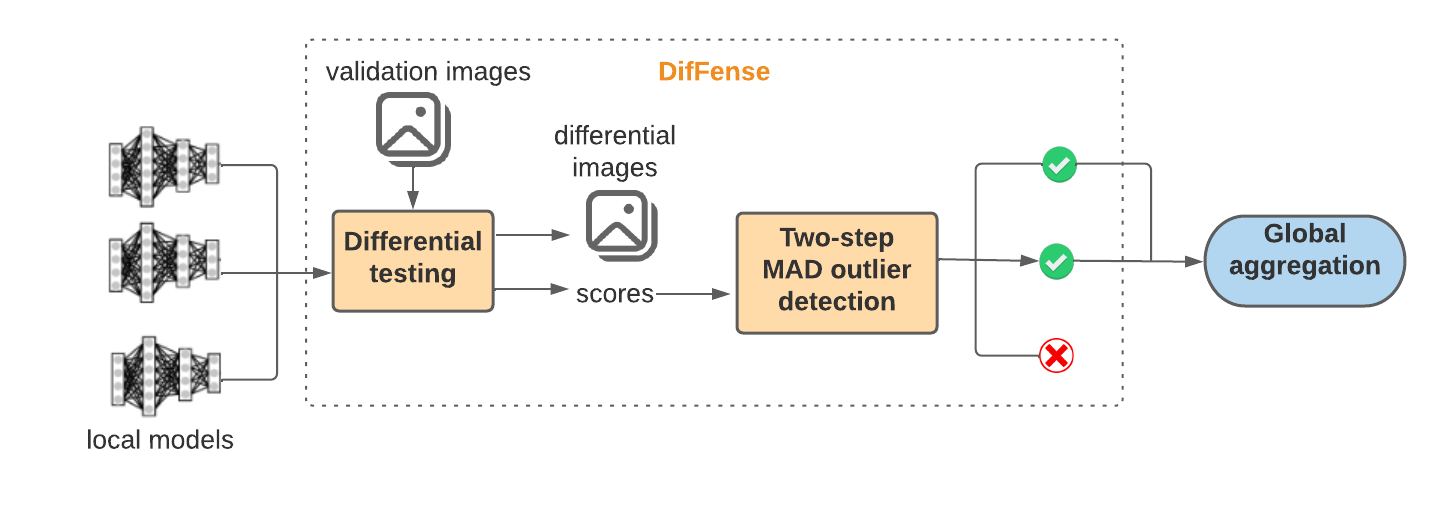}
\vspace{-0.2em}
\caption{Global flow of \sys{} for detecting backdoor attacks in the federated learning system.}
\label{fig_flow}
\end{figure}

%\vspace{-0.6em}
Our system consists of the following steps as in Figure \ref{fig_flow}:
\begin{enumerate}
    % \vspace{-1.2em}
    \item Local training: Local agents are selected to download and train the current global model independently;
    % \vspace{-0.5em}
    \item Selected agents submit their model updates to server;
    % \vspace{-0.5em}
    \item Defense mechanism: 
    \begin{enumerate}
        % \vspace{-0.5em}
        \item Differential input generation: inputs validation images and outputs differential inputs and scores (to be defined in \ref{two_step_mad_outlier_detection}).
        \item Two-step MAD outlier detection: inputs scores and outputs a list of outliers
    \end{enumerate}
    % \vspace{-0.8em}
    \item Server aggregation: exclude the outliers and update the global model by the average of inliers' updates
\end{enumerate}

\vspace{-0.5em}

%\vspace{-1em}
% \subsection{Differential input and score generation}
\subsection{Differential Testing}
\label{differential_input_score_generation}
Given validation images, we process differential inputs in batches and compute the score per class for efficiency. As the result, at the end of this stage, we have a batch of differential images and scores for local agents per image class.

\textbf{Differential input generation.} 
Given $K$ agents and $C$ image classes, we denote $\textbf{X}$ as a batch of images labeled as $c \in [C]$. We modify $\textbf{X}$ iteratively where each iteration consists for the following steps.
\begin{enumerate} 
\item Each local model $i \in [K]$ outputs a prediction on each image $\textbf{x} \in \textbf{X}$ as $p^{i}(\textbf{x}) \in \mathbb{R}^C$. In general, the prediction is a softmax function.
\item Take the average of the model's predictions over the batch to get a vector $\textbf{p}^i \in \mathbb{R}^C$.
\item Optionally reduce the dimension of $\textbf{p}^i$ by PCA projection to get $\textbf{q}^i \in \mathbb{R}^d$ where $d < C$. $d$ can be determined by computing and thresholding the explained variance ratio. The purpose of this step is to reduce the dimensionality and hence, make the clustering easier in the next step.
\item Group the PCA projected model predictions into two clusters, $G_1$ and $G_2$ by running k-means clustering with 2 components. Without loss of generality, $G_1$ is the ``majority cluster" where $|G_1| >= |G_2|$.
\item Given the center vectors of the two clusters, $\boldsymbol{\mu}_1$ and $\boldsymbol{\mu}_2$, run gradient ascent on the batch $\textbf{X}$ by taking the gradient of the objective function $||\boldsymbol{\mu}_1 - \boldsymbol{\mu}_2||_2$ with step size $s$.
\end{enumerate}

\vspace{-0.5em}
Note that there are two hyperparameters, the step size in gradient ascent and the number of iterations. They can be tuned according to the dataset and the model architecture.

% \vspace{-0.5em}
\textbf{Score generation.} We generate scores or ``test statistics" used for hypothesis testing in the next stage. To help determining the confidence level of whether a local agent is adversary, we interpret score as the deviation from the majority cluster center. Instead of calculating a distinct score for each local model, we output two scores - one for the ``majority" and the ``minority" cluster respectively - per image class. This will make the process more scalable to the number of participant models. The goal is to understand the distantness between the two clusters given these scores. 

Models in the majority cluster will be assigned twice the norm of the majority cluster's standard deviation. Assuming normal distribution, $95\%$ of points are within two standard deviations from the mean. It follows from this assumption that two standard deviation is a decent approximation of how far away a local model prediction is from the center for the majority cluster. On the other hand, models in the minority cluster will be assigned the norm of the distance between the two cluster centers. Therefore, the smaller the standard deviation norm of the majority cluster (the more concentrated the majority cluster is) and the greater the distance between the cluster centers are, the greater the difference between the two scores will be and the more likely the minority cluster will be detected as outliers in the ensuing hypothesis testing.

% \vspace{-0.5em}
\textbf{Complexity.} We note that the computational complexity of this stage primarily depends on the number of distinct classes of validation images along with the number of images and the number of iterations.

\begin{algorithm}[h!]
  %\SetKwData{Left}{left}\SetKwData{This}{this}\SetKwData{Up}{up}
  \SetKwFunction{KMeansCluster}{KMeansCluster}
  \SetKwFunction{PCA}{PCA}
  \SetKwInOut{Input}{Input}\SetKwInOut{Output}{Output}
 
  \Input{ seeds $\leftarrow$ original test images \\
  dnns $\leftarrow$ local DNNs \\
  classes $\leftarrow$ image classes \\
  s $\leftarrow$ step size in gradient ascent \\
  t $\leftarrow$ number of iterations \\
  d $\leftarrow$ number of PCA components %\\
  }
  \Output{ gens $\leftarrow$ generated differential images \\
  scores $\leftarrow$ scores of local DNNs
  }
  \BlankLine
  gens := empty set
  
  scores := empty dictionary
  
  \For{$c \in $ classes}{
    \tcp{X is a batch of images belonging to class $c$}
    X $= seeds \in c$  
    
    \For{$i\leftarrow 1$ \KwTo $t$}{
    preds := empty set \\
    \For{dnn $\in$ dnns}{
    \tcp{average of predictions on images of class $c$}
    preds.add((dnn.predict(x) for x in X).mean)
    } 

    predsPca = \PCA(preds, d)
    
    \tcp{$G_1$ is the majority cluster ($|G_1| \geq |G_2|$)}
    $G_1$, $G_2$ = \KMeansCluster(predsPca, 2)  
    
    obj = $||$$G_1$.mean - $G_2$.mean$||$ 
    
    grad = $\partial$obj / $\partial$X
    
    {X} = {X} + s $\cdot$ grad \\
    }
    gens.add({X}) \\
    
    \For{dnn $\in G_1$}{ \tcp{score of the majority cluster is twice its standard deviation norm }
        scores[dnn].add($2*||G_1$.std$||$)
    }
    \For{dnn $\in G_2$}{ \tcp{score of the minority cluster is clusters' mean difference norm}
        scores[dnn].add($||G_1$.mean - $G_2$.mean$||$)
    }
    } 
  return gens, scores
\caption{Differential input and score generation}\label{algo_diffinputgeneration}
\end{algorithm}%\DecMargin{1em}

\vspace{-0.5em}
\subsection{Two-step MAD Outlier Detection}
\label{two_step_mad_outlier_detection}
%\vspace{-0.5em}
Given a matrix of the scores concatenated across classes from \ref{differential_input_score_generation}, $\textbf{P} \in \mathbb{R}^{C \times K}$, we run a statistical test to identify outliers, suspected adversaries.

\vspace{-0.3em}
First, we take the row-wise average of $\textbf{P}$ to get the mean score vector $\textbf{p} \in \mathbb{R}^K$. This makes our score data one-dimensional. Median Absolution Deviation (MAD) outlier detection method~\cite{mad} makes a good candidate as a detection method whose measure of distance is robust to outliers in one-dimensional data. 

\vspace{-0.3em}
As most models belong to the majority cluster and few in the minority have higher score values, we expect our data to be asymmetric with a long tail on the right side. This asymmetric distribution inspires us to use double MAD~\cite{doubleMAD} where two different MADs are used as separate denominators for points below and above the median. 

We observe a few caveats in our problem. Especially when validation images are of only few classes, we expect more than $50\%$ of data to have identical values, incurring ``zero MAD case". Furthermore, assuming that only a small fraction of adversaries participate in the server aggregation, there is a higher chance for models to be adversaries if there are fewer of them. For instance, models in the minority cluster are more likely to be adversaries if they take $10\%$ of the total participants than $40\%$ even with the same score. 
To sum up, we want a double MAD outlier detection method with the following properties:
\begin{itemize}
    % \vspace{-1.3em}
    \item Tolerant of points whose deviations are smaller than the MAD;
    % \vspace{-0.6em}
    \item For points with deviations greater than the MAD, takes into account their ratio and absolute deviation value from the median;
    % \vspace{-0.6em}
    \item Robust to the zero MAD case.
\end{itemize}
% \vspace{-0.8em}
To satisfy these properties, we propose a two-step MAD-based outlier detection using modified double MAD. In this method, we introduce the second MAD $mad_2$ besides the original MAD denoted $mad_1$.
Let $m_p = med(\textbf{p})$, we compute the absolute deviation from the median vector as $\textbf{q}$ where $\textbf{q}(i) = |\textbf{p}(i) - m_p|$ for $i \in [K]$. Then the original MAD $mad_1 = \med(\textbf{q})$ is obtained. 
We compute the second MAD $mad_2$ as the weighted average of $mad_1$ and the median of absolute deviations greater than $mad_1$: 
\begin{align*}
    mad_2 &= \frac{|\{ i | q_i \leq mad_1 \}|}{K} mad_1 \\ &+ \frac{|\{ i | q_i > mad_1 \}|}{K} med(\{ q_i | q_i > mad_1 \}).
\end{align*}
% $mad_2 = \frac{|\{ i | q_i \leq mad_1 \}|}{K} mad_1 + \frac{|\{ i | q_i > mad_1 \}|}{K} med(\{ q_i | q_i > mad_1 \})$.
where $mad_1$ is the denominator for points deviated less than $mad_1$. Utilizing the conventional outlier cutoff value of $2$ or $3$, these points will all be determined as inliers. For points deviated more than $mad_1$, we use $mad_2$ as the denominator. Hence $\textbf{r} \in \mathbb{R}^K$ is defined for $i \in [K]$ as
$$
\textbf{r}(i) = 
\begin{cases}
\frac{\textbf{q}(i)}{mad_1} & \text{ if } \textbf{q}(i) \leq mad_1 \\
\frac{\textbf{q}(i)}{mad_2} & \text{ if } \textbf{q}(i) > mad_1
\end{cases}
$$
Finally, given a pre-defined cutoff threshold $m$ (conventionally $2$ or $3$), we identify the set of outliers as $\{ i | \textbf{r}(i) > m \}$.

\begin{algorithm}[h!]
  %\SetKwData{Left}{left}\SetKwData{This}{this}\SetKwData{Up}{up}
  %\SetKwFunction{KMeansCluster}{KMeansCluster}
  \SetKwInOut{Input}{Input}\SetKwInOut{Output}{Output}
 
  \Input{ dnns $\leftarrow$ local DNNs \\
  scores $\leftarrow$ outlier scores of local DNNs \\
  %k$_1$ $\leftarrow$ weight on mad$_1$ \\
  %k$_2$ $\leftarrow$ weight on mad$_2$ \\
  m $\leftarrow$ threshold
  }
  \Output{ outliers $\leftarrow$ local DNNs identified as outliers
  }
  \BlankLine
  meanScores := empty dictionary
  
  diffs := empty dictionary 
  
  madDiffs := empty dictionary
  
  outliers := empty set
  
  \For{dnn $\in$ dnns}{
  meanScores[dnn] = scores[dnn].mean
  }
  med = meanScores.values.median \\
  
  \For{dnn $\in$ dnns}{
  diffs[dnn] = $|$meanScores[dnn] - med$|$
  }
  diffVals = diffs.values
  
  mad$_1$ = diffVals.median
  
  \tcp{ratio of models with difference greater than the median}
  mad$_2$ratio = (diffVals $>$ mad$_1$).count/dnns.count 
  
  \tcp{mad$_2$ is the weighted average of mad$_1$ and the median of those greater than mad$_1$}
  mad$_2$ = mad$_2$ratio $*$ diffVals[diffVals $>$ mad$_1$].median + (1 - mad$_2$ratio) $*$ mad$_1$ 
 
  \For{dnn $\in$ dnns}{
  \If{diffs[dnn] $\leq$ mad$_1$} {
  \If{mad$_1 = 0$} {
    madDiffs[dnn] = 0
  } \Else{
    madDiffs[dnn] = diffs[dnn]/mad$_1$
  }
  } \Else{
    madDiffs[dnn] = diffs[dnn]/mad$_2$
  }
  }
  \For{dnn $\in$ dnns}{
  \If{madDiffs[dnn] $>$ m} {
  outliers.add(dnn)
  }
  }
return outliers 
\caption{Two-step MAD outlier detection}\label{algo_madoutlierdetection}
\end{algorithm}%\DecMargin{1em}
%%%%%%%%%%%%%%%%%%%%%%%%%%%
% \vspace{-0.6em}
\section{Experiments}
\label{experiment}
% \vspace{-0.2em}
\noindent \textbf{FL aggregation rules.} We evaluate the performance of \sys{} detection-based defense scheme and compare it against FL systems equipped with three different update rules. We assume that at global epoch $t$, $K$ agents out of $N$ total agents are selected for aggregation. We describe the FL aggregation rule of each method below.

% \vspace{-0.2em}
{\tikz\draw[black,fill=black] (-0.5em,-0.5em) rectangle (-0.2em,-0.2em);} \textbf{Federated averaging (FedAvg).} 
This is the conventional global model update rule without any defense, thus is used as the baseline in our experiments.  
The server updates the weights of the global model as:
\begin{equation}
    \label{eq:FedAvg}
     G^{t+1} = G^t + \frac{\eta}{N} \sum_{i=1}^{K} (L_i^{t+1} - G^t),
\end{equation}
% $$ G^{t+1} = G^t + \frac{\eta}{K} \sum_{k} (W_k^{t+1} - G^t),$$
where $\eta$ is the global learning rate and $L_i^{t+1}$ is the model parameter of the $i$-th participating local agent at the $(t+1)$-th epoch.
Throughout the experiments, the global learning rate is set to $\eta = \frac{1}{\beta}$ so that we take the average of only the participating models' weight updates.

\vspace{0.2em}
{\tikz\draw[black,fill=black] (-0.5em,-0.5em) rectangle (-0.2em,-0.2em);} \textbf{Multi-Krum (Krum).}
Assuming that there are $b$ adversaries in the FL system, Krum requires the number of participants to satisfy $K \geq 2b + 3$~\cite{krum}. 
For the $i$-th local update $\delta_i^{t+1} = L_i^{t+1} - G^t$, Krum select a set of its $(K-b-2)$ closest local updates $N_i$ and add up the pairwise distances to obtain a score: $S(\delta_i^{t+1}) = \sum_{\delta \in N_i} || \delta_i^{t+1} - \delta||_p$. 
Krum chooses $l$ local updates with the smallest distance scores $S(\delta_i^{t+1})$ as $S_{t+1}$ (such that $|S_{t+1}| = l$ where $1 \leq l \leq K$) at each iteration and takes their average $\delta_{krum}^{t+1} = \avg_{i \in S_{t+1}} \delta_i^{t+1}$ and uses it to update the global model: 
\vspace{-0.2em}
\begin{equation}
    \label{eq:krum}
    G^{t+1} = G^t +  \delta^{t+1}_{krum}.
    \vspace{-0.2em}
\end{equation}
In our experiment, we assume that Multi-Krum knows the number of adversaries, $b$ in advance and chooses $l = K-b$ local updates for server aggregation.

% \vspace{-0.3em}
{\tikz\draw[black,fill=black] (-0.5em,-0.5em) rectangle (-0.2em,-0.2em);} \textbf{Coordinate-wise median (Coomed).}
The paper~\cite{coomed} shows that taking the coordinate-wise median can achieve the order-optimal statistical error rate in the case of a strongly convex loss function. 
The server updates the shared model's weights as: 
\begin{equation}
    \label{eq:CooMed}
    G^{t+1} = G^t  + \bar{\delta}^{t+1}, 
\end{equation}
% $$ G^{t+1} = G^t  + \bar{\delta}^t$$ 
where $\bar{\delta}^{t+1} := coomed(\delta_i^{t+1} |_{i=1, \ldots K})$ with the $j^{th}$ coordinate $\bar{\delta}^{t+1} (j) = median(\delta_i^{t+1}(j)|_{i=1, \ldots K})$

\vspace{0.3em}
\noindent \textbf{Evaluation metrics.} 
We use two evaluation metrics, global accuracy and backdoor accuracy throughout our experiments. 
Global accuracy is the accuracy of the shared model on the main task (i.e., the proportion of data examples that are correctly classified). Backdoor accuracy is the shared model's accuracy on the backdoor task, which is essentially the attack success rate. 
We measure the final global accuracy after convergence and the average backdoor accuracy during the global training process.

To assess \sys{}'s backdoor detection performance, we measure the false positive rates and false negative rates and report the average values over the entire training process.

When evaluating four aggregation methods, FedAvg, Krum, Coomed, and \sys{}, we identify the epoch that the global model starts converging and compare the maximum backdoor accuracy after that epoch. 

%-------------------------------------------------------------------------
% \vspace{-0.3em}
\noindent \textbf{Experimental setup.} 
We detail the experimental setup and defense configurations of \sys{} below. 
% We detail the experimental setup of \sys{} in terms of attack scenarios and defense configurations below. 

% \vspace{-0.3em}
{\tikz\draw[black,fill=black] (-0.5em,-0.5em) rectangle (-0.2em,-0.2em);} \textbf{Dataset and DNN models.} 
We conduct our experiments on two image datasets, Fashion MNIST and CIFAR10. 
Fashion MNIST~\cite{fmnist} is a dataset of Zalando's black and white article images. It includes $60,000$ training images, $10,000$ test images and $10$ classes. %A data sample has a dimension of $28\times 28 \times 1$. 
We use a custom three-layer Convolutional Neural Network (CNN) that consists of 2 convolutional layers and a fully connected output layer.
The CIFAR10 dataset~\cite{cifar10} includes RGB images of 10 classes. %dimension $32\times32\times3$ in 10 classes.  
The training set contains $50,000$ samples and the test set has $10,000$ samples. We use a lightweight ResNet-18 CNN model{~\cite{resnet}} with $2.7$ million parameters as the model architecture. 

% \vspace{-0.3em}
{\tikz\draw[black,fill=black] (-0.5em,-0.5em) rectangle (-0.2em,-0.2em);} \textbf{FL setting.} We assume that there are $N=50$ local clients with the participation ratio of $\beta = 20\%$ (hence, at each global epoch, $m=10$ number of agents will be selected for aggregation). The default number of adversaries in the FL system is set to 1 if not specified otherwise.

% \vspace{-0.3em}
{\tikz\draw[black,fill=black] (-0.5em,-0.5em) rectangle (-0.2em,-0.2em);} \textbf{Backdoor attack setting.} This attack variant includes both pixel-pattern and semantic backdoor attacks. 
In this scenario, we assume the server trains a global model for $150$ iterations on Fashion MNIST and $300$ iterations on CIFAR10 dataset. 
%The global learning rate is set to $\eta = \frac{1}{\beta} = 5$ for both datasets, so that we take the average of only the participating models' weight updates.
The training data is distributed to local agents in a  \textit{non-i.i.d} fashion based on a Dirichlet distribution with hyper-parameter $0.9$~\cite{dirichlet}.
The attackers use the objective function in Equation~(\ref{eq:adv_obj}) with the hyperparameter $\alpha=0.7$ to obtain malicious local updates while regularizing the deviation or the $L_2$ norm distance from the prior global model parameter.
Both benign and adversarial models are trained locally for $3$ epochs with a learning rate of $0.05$, momentum of $0.9$ and decay of $0.0005$.
The weight-scaling factor of the adversary is $\frac{N}{\eta} = 10$ where $N$ is the number of total participants and $\eta$ is the global learning rate. When there are $b > 1$ adversaries at an epoch, we assume that they ``collude" by evenly distributing the scaling factor by $b$. Reducing the scaling factor may help the adversaries avoid being detected by the server.

We detail the configuration of each attack below: 
\begin{itemize}
    % \vspace{-0.6em}
    \item \textit{Pixel-pattern backdoors} require the modification of inputs in a special way during training and testing. 
    In the training stage, we use a white square sticker of size $4$ on the bottom-right corner as the trigger and add it to 20 images of each data batch of size $64$. Following the paper~\cite{backdoor}, we add the Gaussian noise of standard deviation $\sigma = 0.01$ to each backdoor image to facilitate model generalization.
    For testing, backdoor images are generated in the same way.
    \item \textit{Semantic backdoors} induce the model to mis-predict unmodified inputs with certain features. 
      As suggested in the paper{~\cite{backdoor}}, we select images with three features from CIFAR10 dataset as semantic triggers - green cars, cars with racing stripes and cars with vertically striped walls in the background.
    %As suggested in the previous paper{~\cite{backdoor}}, we select images with three features from CIFAR10 dataset as semantic triggers - green cars (30 images), cars with racing stripes (21 images) and cars with vertically striped walls in the background (12 images).
    These chosen images are mislabeled as ``airplane".
    As in pixel-pattern attacks, we add the Gaussian noise of standard deviation $\sigma = 0.01$ to each backdoor image.
    We split the training data so that only adversaries have access to the backdoor images. 
    Out of total 63 backdoor images, 48 images are used for training, and 15 are held out for the test phase. 
    An attacker trains his local model with a batch size of 64 and each batch includes 20 backdoor images. 
    %$1000$ test images are generated by randomly flipping and cropping the held-out images.    
    Backdoor test images, generated by randomly flipping and cropping the held-out images, are used to compute the backdoor accuracy of the global model. 
\end{itemize}
% \vspace{-1.5em}
To sum up, we experiment with three attack scenarios: $(i)$ pixel-pattern attack on Fashion-MNIST, $(ii)$ pixel-pattern attack on CIFAR10 and $(iii)$ semantic attack on CIFAR10.

{\tikz\draw[black,fill=black] (-0.5em,-0.5em) rectangle (-0.2em,-0.2em);} \textbf{\sys{} configuration.}  
The workflow of \sys{} is described in Algorithm~\ref{algo_diffinputgeneration}, \ref{algo_madoutlierdetection}. We set the step size in gradient descent $s = 0.5$, the number of iterations $t = 20$ and the number of PCA components $d = 2$ when generating differential inputs.
The cutoff threshold in the two-step MAD outlier detection is set to $m=3$. 

In \ref{ablation_study}, we analyze \sys{}'s performance against the number and distribution of differential images. In other experiments, we generate 20 differential images chosen randomly with the uniform distribution from 10 classes.

%% --------- 
%\vspace{-0.3em}
\subsection{Results} \label{results}
%\vspace{-0.3em}
We compare the performance between FedAvg (baseline) and \sys{} under pixel-pattern attack in Table \ref{tab_pixelcomparison} and semantic attack in Table \ref{tab_multipleadvs}, given one adversary. We can observe that \sys{}'s global model accuracy is comparable to FedAvg's. 
Regardless of the trigger pattern, \sys{} is more robust to attacks with the average backdoor accuracy of below $2\%$.
\sys{}'s strong FL performance is attributed to its low false positive rate of below $7\%$ and zero false negative rate as shown in Table \ref{tab_fmnistfprfnr1}, \ref{fprfnr_cifar10}.

\begin{figure}[ht!]
\centering
\begin{tabular}{ccc}
\includegraphics[width=0.27\linewidth]{"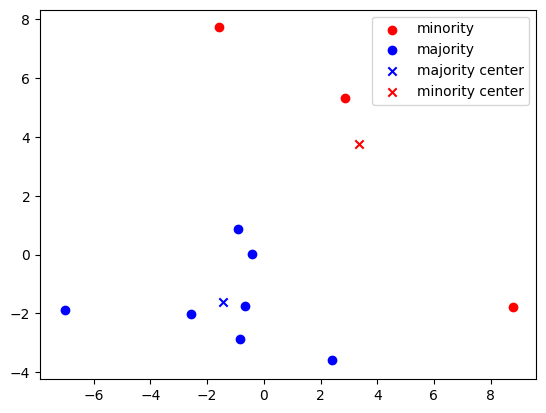"} &
\includegraphics[width=0.3\linewidth]{"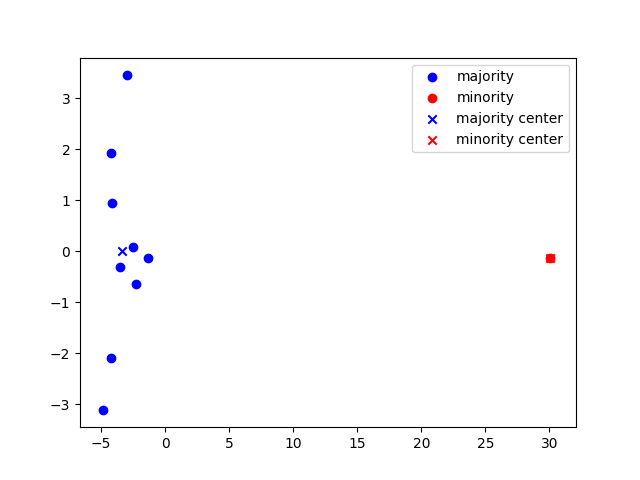"} & 
\includegraphics[width=0.3\linewidth]{"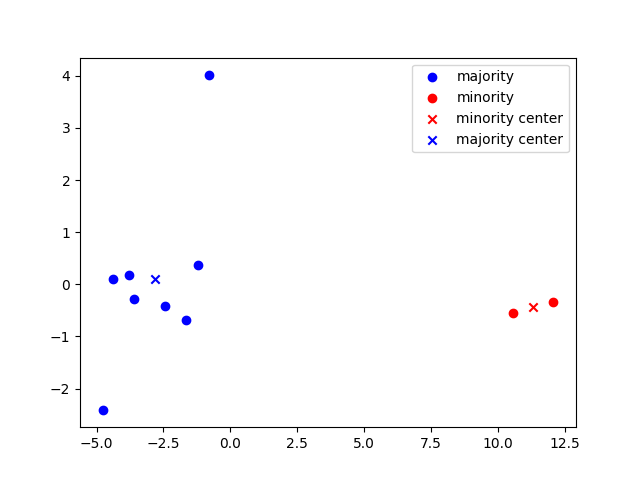"} \\
a) & b) & c)\\
\end{tabular}
% \vspace{-1.2em}
\caption{Distribution of PCA-projected predictions by two clusters a) without, b) with an adversary and c) with two adversaries}
\label{fig_pca}
\end{figure}

\begin{figure}[ht!]
\begin{tabular}{cc}
\includegraphics[width=0.45\linewidth]{"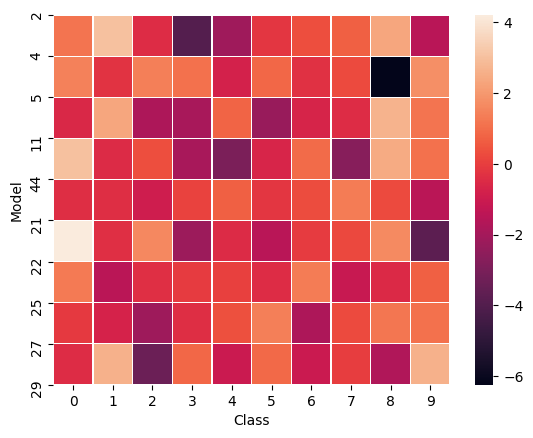"} &
\includegraphics[width=0.45\linewidth]{"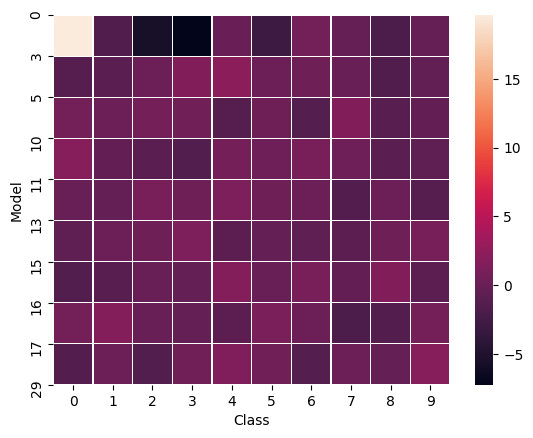"} \\
a) & b)\\
\end{tabular}
% \vspace{-1.2em}
\caption{Average prediction score distribution on differential inputs of ten models at an epoch a) without and b) with an adversary  who targets the label of backdoor images at ``0"}
\label{fig_pred}
\end{figure}

We further observe that differential testing is effective for distinguishing adversaries from benign agents. In Figure \ref{fig_pca}, adversaries' PCA-projected predictions on a differential image are clearly segregated from the benign models' regardless of the number of participant attackers. In particular, as shown in figure \ref{fig_pred}, the average prediction distribution of an adversary on differential inputs stands out, outputting an abnormally high value on the targeted class.

\begin{table}[ht!]
\caption{Performance comparison between the FedAvg and \sys{} against \textbf{pixel-pattern} attacks on \textbf{Fashion-MNIST} after 150 global epochs and \textbf{CIFAR10} after 300 global epochs.}
\vspace{-0.3em}
\begin{center}
\begin{small}
\begin{sc}
\scalebox{1.1}{
\begin{tabular}{c|c|c|c|c|}
\cline{2-5}
& \multicolumn{2}{c|}{\textbf{F-MNIST}} 
& \multicolumn{2}{c|}{\textbf{CIFAR10}} \\ 
\cline{2-5} 
\textbf{}                               & \textbf{FedAvg}    & \textbf{\sys{}}    & \textbf{FedAvg}  & \textbf{\sys{}}  \\ \hline
\multicolumn{1}{|c|}{\textbf{Final GA ($\%$)}} & 82.1                & 84.3                    & 89.62              & 88.87                  \\ \hline
\multicolumn{1}{|c|}{\textbf{Mean BA ($\%$)}}  & 6.15                & 0.68                    & 7.71               & 0.25                  \\ \hline
\end{tabular}
}
\end{sc}
\end{small}
\end{center}
\vskip -0.1in
\label{tab_pixelcomparison}
\end{table}

\begin{table}[ht!]
\caption{Average of false positive rate and false negative rate of \sys{} against \textbf{pixel-pattern} attacks on \textbf{Fashion-MNIST} and \textbf{CIFAR10}.}
%\vskip 0.15in
\begin{center}
\begin{small}
\begin{sc}
\scalebox{1.1}{
\begin{tabular}{c|c|c|}
\cline{2-3}
\textbf{}    & \textbf{F-MNIST} & \textbf{CIFAR10} \\ \hline
\multicolumn{1}{|c|}{\textbf{Mean FPR ($\%$)}} & 5    & 5   \\ \hline
\multicolumn{1}{|c|}{\textbf{Mean FNR ($\%$)}} & 0    & 0    \\ \hline
\end{tabular}
}
\end{sc}
\end{small}
\end{center}
\vskip -0.1in
\label{tab_fmnistfprfnr1}
\end{table}

\vspace{-0.5em}

%\vspace{-1em}

\subsection{Comparative Analysis against Multiple Adversaries} \label{comp_anlaysis}
%\vspace{-0.3em}
We evaluate the performance of \sys{} against FedAvg and other defense mechanisms, Multi-Krum and CooMed, given multiple adversaries ranging from 1 to 3. We run experiments on CIFAR10 dataset under semantic attack. The higher the number of adversaries is, the more frequently an adversary may be chosen for the server aggregation. Even when all the three adversaries are selected at an epoch, $K \geq 2b+3$ with the number of participant models, $K = 10$ and the number of adversaries, $b = 3$. Hence, our attack setting satisfies the Krum's requirement.

%% ---- sensitivity to num of advs ----# 
\begin{table*}[t!]
\caption{Performance comparison between FedAvg, Multi-Krum (denoted as Krum), CooMedian and \sys{} against \textbf{semantic} attack on \textbf{CIFAR10} dataset with a different number of adversaries.}
% \vspace{-0.9em}
% \resizebox{\columnwidth}{!}{
\begin{center}
\begin{small}
\begin{sc}
\resizebox{\columnwidth}{!}{
\begin{tabular}{|c|cccc|cccc|cccc|}
\hline
Number of adversaries & \multicolumn{4}{c|}{\textbf{1}}                                                                                                         & \multicolumn{4}{c|}{\textbf{2}}                                                                                                         & \multicolumn{4}{c|}{\textbf{3}}                                                                              \\ \hline
Aggregation method    & \multicolumn{1}{c|}{\textbf{FedAvg}} & \multicolumn{1}{c|}{\textbf{Krum}} & \multicolumn{1}{c|}{\textbf{CooMed}} & \textbf{\sys{}} & \multicolumn{1}{c|}{\textbf{FedAvg}} & \multicolumn{1}{c|}{\textbf{Krum}} & \multicolumn{1}{c|}{\textbf{CooMed}} & \textbf{\sys{}} & \multicolumn{1}{c|}{\textbf{FedAvg}} & \multicolumn{1}{c|}{\textbf{Krum}} & \multicolumn{1}{c|}{\textbf{CooMed}} & \textbf{\sys{}} \\ \hline
\textbf{Final GA}              & \multicolumn{1}{c|}{89.06}           & \multicolumn{1}{c|}{90.04}         & \multicolumn{1}{c|}{85.91}           & 89.41                & \multicolumn{1}{c|}{89.7}            & \multicolumn{1}{c|}{89.87}         & \multicolumn{1}{c|}{88.82}           & 84.1                & \multicolumn{1}{c|}{89.18}           & \multicolumn{1}{c|}{87.94}         & \multicolumn{1}{c|}{88.89}            & 89.77 \\ \hline
\textbf{Max BA}                & \multicolumn{1}{c|}{100}             & \multicolumn{1}{c|}{34.8}          & \multicolumn{1}{c|}{45.9}           & 10.5                  & \multicolumn{1}{c|}{99.9}             & \multicolumn{1}{c|}{96.8}            & \multicolumn{1}{c|}{94.5}            & 11.9                  & \multicolumn{1}{c|}{100}             & \multicolumn{1}{c|}{98.8}          & \multicolumn{1}{c|}{96.8}             & 15.5                  \\ \hline
\textbf{Mean BA}               & \multicolumn{1}{c|}{14.85}           & \multicolumn{1}{c|}{2.84}          & \multicolumn{1}{c|}{4.4}             & 0.41                 & \multicolumn{1}{c|}{17.86}           & \multicolumn{1}{c|}{2.07}          & \multicolumn{1}{c|}{9.69}           & 1.15                 & \multicolumn{1}{c|}{24.8}           & \multicolumn{1}{c|}{3.6}          & \multicolumn{1}{c|}{4.29}            & 1.94                 \\ \hline
\end{tabular}
}
\end{sc}
\end{small}
\end{center}
%\vskip -0.1in
\vspace{-1em}
%}
\label{tab_multipleadvs}
%\end{table}
\end{table*}

We can observe in Table \ref{tab_multipleadvs}, \ref{fprfnr_cifar10} that \sys{} obtains a lower average backdoor accuracy than other methods with consistently zero false negative rate while achieving a global accuracy comparable to the baseline with below $7\%$ of false positive rate. Compared to Multi-Krum, \sys{} successfully detects all the adversaries while yielding a lower false positive rate, regardless of the number of adversaries.

We also compute the maximum backdoor accuracy of each method after the global model reaches the test accuracy of $85\%$ in Table \ref{tab_multipleadvs}. 
\sys{}' backdoor accuracy is consistently lower than $16\%$ while other methods' reach $95\%$ with multiple adversaries. To sum up, \sys{} steadily presents a stronger performance than FedAvg, Multi-Krum and CooMed given a various number of adversaries.

% \vspace{-1em}
\begin{table}[h!]
\caption{Mean false positive rate and false negative rate of \sys{} and Multi-Krum over 300 rounds given different numbers of adversaries under \textbf{semantic} attack on \textbf{CIFAR10} dataset.}
% \vspace{-1em}
%\centering
\begin{center}
\begin{small}
\begin{sc}
\resizebox{0.7\columnwidth}{!}{
\begin{tabular}{|c|cc|cc|cc|}
\hline
Number of adversaries                    & \multicolumn{2}{c|}{\textbf{1}}                                                & \multicolumn{2}{c|}{\textbf{2}}                                                & \multicolumn{2}{c|}{\textbf{3}}                                                \\ \hline
\multicolumn{1}{|l|}{Aggregation method} & \multicolumn{1}{l|}{\textbf{Krum}} & \multicolumn{1}{l|}{\textbf{\sys{}}} & \multicolumn{1}{l|}{\textbf{Krum}} & \multicolumn{1}{l|}{\textbf{\sys{}}} & \multicolumn{1}{l|}{\textbf{Krum}} & \multicolumn{1}{l|}{\textbf{\sys{}}} \\ \hline
\textbf{Mean FPR}                        & \multicolumn{1}{c|}{8.23}          & 6.7                                       & \multicolumn{1}{c|}{16}          & 4.16                                      & \multicolumn{1}{c|}{27.17}          & 3.54                                      \\ \hline
\textbf{Mean FNR}                        & \multicolumn{1}{c|}{0}             & 0                                         & \multicolumn{1}{c|}{0.8}          & 0                                         & \multicolumn{1}{c|}{0.48}         & 0                                         \\ \hline
\end{tabular}
}
\end{sc}
\end{small}
\end{center}
\vskip -0.1in
\label{fprfnr_cifar10}
\end{table}

\vspace{-1em}
\begin{figure}[ht!]
\begin{tabular}{ccc}
\includegraphics[width=0.46\linewidth]{"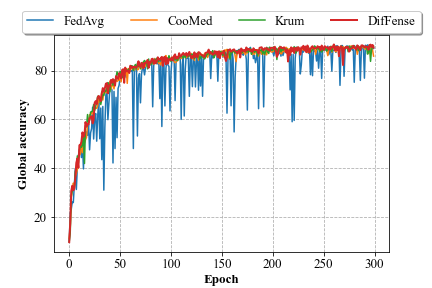"} &
\includegraphics[width=0.46\linewidth]{"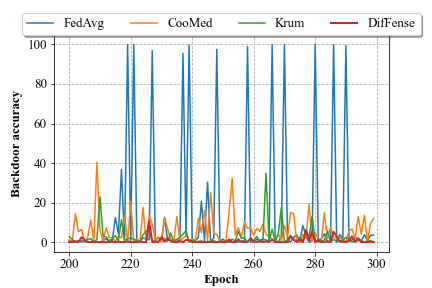"} \\
a) & b) \\
\end{tabular}
\vspace{-0.5em}
\caption{Progression plot of a) the global accuracy throughout the entire training process and b) the backdoor accuracy during the last 100 epochs under semantic attack on CIFAR10 given 1 adversary}
\end{figure}

\vspace{-0.5em}
\subsection{Ablation Study} \label{ablation_study}
\noindent \textbf{Number of differential images.}
We run a sensitivity analysis to investigate whether \sys{}'s effectiveness is correlated with the number of generated differential images. We can observe that the final global accuracy is consistently around $89\%$ while the average backdoor accuracy is below $4\%$. The average false positive rate is also comparable, ranging from $5\%$ to $8\%$ and the average false negative rate is zero.

% \vspace{-1em}
\begin{table}[h!]
\caption{Performance analysis of \sys{} with respect to the number of differential images generated with 300 epochs of global training on \textbf{CIFAR10} dataset under \textbf{semantic} attack.}
%\vspace{0.3em}
\begin{center}
\begin{small}
\begin{sc}
\resizebox{0.5\columnwidth}{!}{
\begin{tabular}{|c|c|c|c|}
\hline
\# Differential Inputs & \textbf{20} & \textbf{50} & \textbf{100} \\ \hline
\textbf{Final GA}      & 89.41       & 89.93        & 89.15        \\ \hline
\textbf{Mean BA}       & 0.41        & 2.49        & 3.14         \\ \hline
\textbf{Mean FPR}      & 6.7        & 7.73        & 5.47         \\ \hline
\textbf{Mean FNR}      & 0           & 0           & 0            \\ \hline
\end{tabular}
}
\end{sc}
\end{small}
\end{center}
\vskip -0.1in
\label{tab_diffimages}
\end{table}

%\vspace{0.5em}
\noindent \textbf{Distribution of differential images.} Fixing the number of differential images as $20$, we differentiate the number of distinct classes that the images belong to. We assume that differential images are \textit{i.i.d} distributed across the given classes. We can observe that the average false positive rate decreases as more classes are used while false negative rate is flat zero. However, even with one class, \sys{} achieves a false positive rate of below $9\%$ and the final global accuracy comparable to the one whose differential images are from all the classes. This alleviates the burden of securing validation images distributed across a number of classes. Also, using images from only few classes can reduce the runtime as the computational complexity of Algorithm \ref{algo_diffinputgeneration} is affected by the number of image classes. 

%\begin{table}[h!]
\vspace{-1.5em}
\begin{table}
\caption{Performance analysis of the \sys{} with respect to the number of distinct classes that 20 differential images belong to, generated with 300 epochs of global training on \textbf{CIFAR10} dataset under \textbf{semantic} attack.}
\vspace{-0.1em}
\begin{center}
\begin{small}
\begin{sc}
\resizebox{0.7\columnwidth}{!}{
\begin{tabular}{|c|c|c|c|c|}
\hline
\# differential input classes & \textbf{1} & \textbf{2} & \textbf{5} & \textbf{10} \\ \hline
\textbf{Final GA}             & 89.71      & 89.86      & 89.51      & 89.54       \\ \hline
\textbf{Mean BA}              & 2.34       & 3.03       & 3.23       & 5.34        \\ \hline
\textbf{Mean FPR}             & 8.87       & 8.93       & 7.37       & 4.43        \\ \hline
\textbf{Mean FNR}             & 0          & 0          & 0          & 0           \\ \hline
\end{tabular}
}
% \resizebox{0.9\columnwidth}{!}{
% \begin{tabular}{|c|c|c|c|c|}
% \hline
% \# differential input classes & \textbf{1} & \textbf{2} & \textbf{5} & \textbf{10} \\ \hline
% \textbf{Final GA}             & 90.21      & 87.07      & 87.61      & 87.51       \\ \hline
% \textbf{Mean BA}              & 4.92       & 1.16       & 1.07       & 0.65        \\ \hline
% \textbf{Mean FPR}             & 6.5        & 9.04       & 6.56       & 5.22        \\ \hline
% \textbf{Mean FNR}             & 0          & 0          & 0          & 0           \\ \hline
% \end{tabular}
% }
\end{sc}
\end{small}
\end{center}
\vskip -0.1in
\label{tab_diffclasses}
\end{table}

\vspace{1.2em}
\noindent \textbf{Two-step MAD outlier detection.} We corroborate the effectiveness of two-step MAD outlier detection compared to the original single and double MAD variants. 
% outlier detection methods.
Single MAD suffers from a high false positive rate while double MAD from a high false negative rate ($100\%$). Two-step MAD outlier detection achieves the zero false negative rate while suppressing the false positive rate to below $7\%$.

% \vspace{-1em}
\begin{table}[h!]
\caption{\sys{}'s performance based on different MAD outlier detection methods on \textbf{CIFAR10} dataset under \textbf{semantic} attack.}
\vspace{-0.1em}
\begin{center}
\begin{small}
\begin{sc}
\resizebox{0.7\columnwidth}{!}{
\begin{tabular}{|c|c|c|c|}
\hline
Outlier Detection & \textbf{Two-step MAD} & \textbf{Double MAD} & \textbf{Single MAD} \\ \hline
\textbf{Final GA} & 89.41                 & 89.21               & 88.31        \\ \hline
\textbf{Mean BA}  & 0.41                  & 14.34               & 6.09         \\ \hline
\textbf{Mean FPR} & 6.7                   & 1.17                & 23.17        \\ \hline
\textbf{Mean FNR} & 0                     & 100                 & 0            \\ \hline
\end{tabular}
}
\end{sc}
\end{small}
\end{center}
\vskip -0.1in
\label{tab_mad}
\end{table}

\vspace{-0.7em}
\section{Conclusion}
\label{conclusion}
\vspace{-0.3em}
In this work, we design and experiment \sys{}, an automated defense framework in FL based on differential testing and two-step MAD outlier detection. 
Our method addresses limitations of the prior defense mechanisms by requiring minimal knowledge of the attack setting. 
We also empirically show that \sys{} yields a backdoor accuracy lower than other aggregation methods while achieving a global accuracy comparable to the baseline, FedAvg. \sys{} successfully detects and excludes any number of adversaries while suppressing a false positive rate below 10$\%$.
The performance of \sys{} is steadily strong regardless of the number and distribution of differential images.

%\begin{table}[ht!]

\Urlmuskip=0mu plus 1mu\relax
\bibliography{refs}
\bibliographystyle{icml2022}

\end{document}